\documentstyle[11pt,newpasp,twoside,epsf]{article}
\def\edcomment#1{\iffalse\marginpar{\raggedright\sl#1\/}\else\relax\fi}
\reversemarginpar

\begin{document}
\title{ Mass of BL Lacs from the 
        Velocity Dispersion of the Host Galaxy}
 \author{A. Treves}
\affil{$^1$ Universit\`a dell'Insubria, via Valleggio 11, 20126 Como, I
\\E-mail  aldo.treves@uninsubria.it}
\author{N. Carangelo$^{1,4}$, R. Falomo$^2$, J. Kotilainen$^3$}
\affil{$^2$ Osservatorio Astronomico di Padova;
$^3$ Tuorla Observatory;\\
$^4$ Universit\`a di Milano-Bicocca}

\setcounter{page}{111}
\index{Author, I.}
\index{Co-Author, I.}

\section{Introduction}

It has been recently shown that in nearby early type galaxies the
central velocity dispersion $\sigma$ correlates rather tightly with the
central black hole mass M$_{BH}$ (Gebhardt et al. 2000, 
Ferrarese and Merritt 2000).
Assuming the M$_{BH}$--$\sigma$ relation is valid also for 
ellipticals hosting active nuclei  
it becomes possible to estimate their BH mass  through
the measurement of $\sigma$. This is complementary to the determination
of M$_{BH}$ through the luminosity of the bulge of the galaxy, basing on the
M$_{BH}$--L$_{bulge}$ dependence  (e.g. Magorrian et al. 1998; 
Kormendy \& Gebhardt 2001).
In particular these two procedures are a promising way for estimating 
the mass of BL Lacs for which,  due to the lack of  emission lines, 
other methods to derive M$_{BH}$ cannot be applied.

Following the systematic study of host galaxies of 110 BL Lacs
with HST (Urry et al 2000) M$_{BH}$ was estimated for a number of
BL Lacs from the M$_{BH}$--L$_{bulge}$ relation (O'Dowd et al 2002; 
Falomo et al 2002b). 
Barth et al (2002 and this conference) and Falomo et al 2002a
adopted the strategy of measuring $\sigma$, and considered a dozen BL Lacs.

Here we report on  new measurements of the stellar velocity dispersion of 
4 BL Lac objects  for which optical  spectroscopy was   gathered at the ESO 3.6m 
using EFOSC2. As example the spectrum  PKS 0521-36, 
showing the absorption features of the host galaxy together with some nuclear 
emission lines,is reported in Figure 1.
The analysis was performed following that described in Falomo et al 2002a
referring to the observations taken at NOT + ALFOSC.

\section {Results}
The values of $\sigma$ and the corresponding M$_{BH}$ are reported in the Table 
(these include both previous NOT and new ESO data; labelled as N and E, 
respectively). 
In the case of PKS 2201+04 observations from both dataset 
are available and indicate a good agreement. 
We estimate that the typical uncertainty on the BH mass 
is 30\%.
Our values are in general agreement with those reported by 
Barth et al 2002, although there are significant discrepancies 
for specific objects (e.g. Mrk 501 and IZw187).
 In the Table we also report for comparison M$_{BH}$ calculated through the 
host galaxy luminosity (cf Falomo et al 2002a,b).
 
It has been proposed that the
 FWHM of the [OIII] 5007 A line correlates with $\sigma$ from the stars 
(Nelson 2000, Shields et al 2002) making the measurement of this emission line
an important probe for estimate of $\sigma$ in high z sources.  
For two objects (PKS 0521-36 and 3C 371) in our sample we have 
significant OIII emission from which $\sigma_{[OIII]}$ can be derived.
We found $\sigma_{[OIII]}$ = 244 Km/s and $\sigma_{[OIII]}$ = 291 Km/s for PKS 0521-36 and 3C371 respectively,  
in good agreement with the values 
derived from the stellar velocity dispersion.\\

\begin{figure}[hb]
\plotfiddle{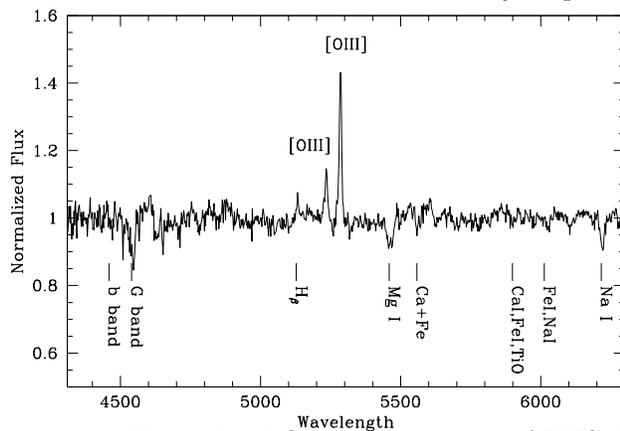}{4cm}{0}{50}{50}{-175}{-175}
\caption{Normalized Optical spectrum of PKS0521-36.}
\end{figure}

\begin{table*}
\centering
{\footnotesize
\begin{tabular}{|l|c|c|c|c|c|c|} \hline
Object & N/E & z & $\sigma_{c}$ & M$_{R}$(host) & Log(M$_{BH})_{\sigma}$ & Log(M$_{BH})_{bulge}$  \\
	& & & Km s$^{-1}$ & & [M$_{\odot}$] & [M$_{\odot}$] \\ \hline
Mrk 421	& N	&0.031&   236   &   -23.15   &  8.53	  & 8.69	      \\
Mrk 180  & N	&0.045&   244   &   -22.84   &	 8.59  &    8.53	       \\
Mrk 501  & N	&0.034&   291   &   -23.91   &	 8.94  &    9.07	      \\
I Zw187  & N	&0.055&   253   &   -22.32   &	 8.66  &    8.27	       \\
3C 371   & N	&0.051&   284   &   -23.89   &	 8.89  &    9.06	      \\
1ES 1959+65 & N	&0.048&   195   &   -23.06   &	 8.15  &    8.64	       \\
PKS 2201+04 & N	&0.027&   160   &   -22.51   &	 7.76  &    8.37      		\\ 
	    & E	&0.027&   165   &   -22.51   &	 7.82  &    8.37	      \\
PKS 0521-36  &E	&0.055&   255   &   -23.24   &	 8.68  &    8.73	       \\
PKS 0548-32  &E	&0.069&   263   &   -23.63   &   8.74  &    8.93	      \\
ApLib  &  E	&0.049&   250  &   -23.48   &	 8.64  &    8.85	       \\ \hline
\end{tabular}					    
}
\end{table*}

\end{document}